\documentclass[twoside,twocolumn,9pt]{article}
\usepackage{extsizes}
\usepackage[super,sort&compress,comma]{natbib} 
\usepackage[version=3]{mhchem}
\usepackage[left=1.5cm, right=1.5cm, top=1.785cm, bottom=2.0cm]{geometry}
\usepackage{balance}
\usepackage{mathptmx}
\usepackage{amsmath,amssymb}
\usepackage{sectsty}
\usepackage{graphicx} 
\usepackage{subcaption}
\usepackage{lastpage}
\usepackage{booktabs}
\usepackage[format=plain,justification=justified,singlelinecheck=false,font={stretch=1.125,small,sf},labelfont=bf,labelsep=space]{caption}
\usepackage{float}
\usepackage{fancyhdr}
\usepackage{fnpos}
\usepackage[english]{babel}
\addto{\captionsenglish}{%
  
}
\usepackage{array}
\usepackage{droidsans}
\usepackage{charter}
\usepackage{lipsum}
\usepackage[T1]{fontenc}
\usepackage[usenames,dvipsnames]{xcolor}
\usepackage{setspace}
\usepackage[compact]{titlesec}
\usepackage{hyperref}
\newcommand{\rulesep}{\unskip\ \vrule\ }

\definecolor{cream}{RGB}{222,217,201}

\begin{document}

\pagestyle{fancy}
\thispagestyle{plain}
\fancypagestyle{plain}{
%%%HEADER%%%
\renewcommand{\headrulewidth}{0pt}
}
%%%END OF HEADER%%%

%%%PAGE SETUP - Please do not change any commands within this section%%%
\makeFNbottom
\makeatletter
\renewcommand\LARGE{\@setfontsize\LARGE{15pt}{17}}
\renewcommand\Large{\@setfontsize\Large{12pt}{14}}
\renewcommand\large{\@setfontsize\large{10pt}{12}}
\renewcommand\footnotesize{\@setfontsize\footnotesize{7pt}{10}}
\makeatother

\renewcommand{\thefootnote}{\fnsymbol{footnote}}
\renewcommand\footnoterule{\vspace*{1pt}% 
\color{cream}\hrule width 3.5in height 0.4pt \color{black}\vspace*{5pt}} 
\setcounter{secnumdepth}{5}

\makeatletter 
\renewcommand\@biblabel[1]{#1}            
\renewcommand\@makefntext[1]% 
{\noindent\makebox[0pt][r]{\@thefnmark\,}#1}
\makeatother 
\renewcommand{\figurename}{\small{Fig.}~}
\sectionfont{\sffamily\Large}
\subsectionfont{\normalsize}
\subsubsectionfont{\bf}
\setstretch{1.125} %In particular, please do not alter this line.
\setlength{\skip\footins}{0.8cm}
\setlength{\footnotesep}{0.25cm}
\setlength{\jot}{10pt}
\titlespacing*{\section}{0pt}{4pt}{4pt}
\titlespacing*{\subsection}{0pt}{15pt}{1pt}
%%%END OF PAGE SETUP%%%

%%%FOOTER%%%
\fancyfoot{}
\fancyfoot[LO,RE]{\vspace{-7.1pt}\includegraphics[height=9pt]{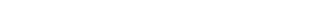}}
\fancyfoot[CO]{\vspace{-7.1pt}\hspace{13.2cm}\includegraphics{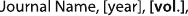}}
\fancyfoot[CE]{\vspace{-7.2pt}\hspace{-14.2cm}\includegraphics{head_foot/RF}}
\fancyfoot[RO]{\footnotesize{\sffamily{1--\pageref{LastPage} ~\textbar  \hspace{2pt}\thepage}}}
\fancyfoot[LE]{\footnotesize{\sffamily{\thepage~\textbar\hspace{3.45cm} 1--\pageref{LastPage}}}}
\fancyhead{}
\renewcommand{\headrulewidth}{0pt} 
\renewcommand{\footrulewidth}{0pt}
\setlength{\arrayrulewidth}{1pt}
\setlength{\columnsep}{6.5mm}
\setlength\bibsep{1pt}
%%%END OF FOOTER%%%

%%%FIGURE SETUP - please do not change any commands within this section%%%
\makeatletter 
\newlength{\figrulesep} 
\setlength{\figrulesep}{0.5\textfloatsep} 

\newcommand{\topfigrule}{\vspace*{-1pt}% 
\noindent{\color{cream}\rule[-\figrulesep]{\columnwidth}{1.5pt}} }

\newcommand{\botfigrule}{\vspace*{-2pt}% 
\noindent{\color{cream}\rule[\figrulesep]{\columnwidth}{1.5pt}} }

\newcommand{\dblfigrule}{\vspace*{-1pt}% 
\noindent{\color{cream}\rule[-\figrulesep]{\textwidth}{1.5pt}} }

\makeatother
\graphicspath{ {Figures/} }

%%%END OF FIGURE SETUP%%%

%%%TITLE, AUTHORS AND ABSTRACT%%%
\twocolumn[
  \begin{@twocolumnfalse}
{\includegraphics[height=30pt]{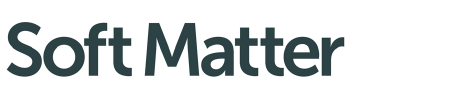}\hfill\raisebox{0pt}[0pt][0pt]{\includegraphics[height=55pt]{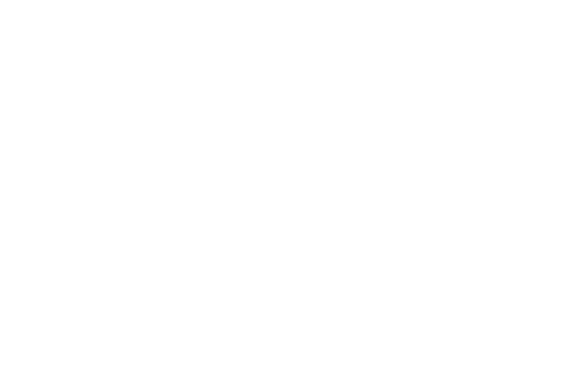}}\\[1ex]
\includegraphics[width=18.5cm]{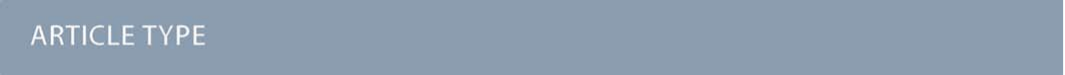}}\par
\vspace{1em}
\sffamily
\begin{tabular}{m{4.5cm} p{13.5cm} }

\includegraphics{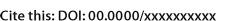} & \noindent\LARGE{\textbf{Braided mixing in confined chiral active matter}} \\
\vspace{0.3cm} & \vspace{0.3cm} \\

 & \noindent\large{Yue Wang\textit{$^{a}$} and Jonas Berx\textit{$^{a\ddag}$}} \\

\includegraphics{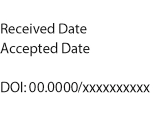} & \noindent\normalsize{Efficient mixing of fluids is essential in many practical applications to achieve homogeneity. For microscopic systems, however, both diffusion and turbulence are ineffective methods to achieve chaotic mixing due to the low Reynolds number, hence either active stirring or inducing turbulence through geometric boundary effects are generally implemented. Here, we study a modified chiral Vicsek model, where active microswimmers act as moving rods, stirring the surrounding substrate. We study the degree of mixing in the patterns formed by interplay between confinement, chiral motion and alignment interactions. This mixing is computed by considering the entanglement of spacetime trajectories of the particles, which forms a braid. Optimising the finite-time braiding exponent of this braid then yields a set of constituent parameters of the system, showing that a pattern consisting of a local stable vortex droplet and an ordered oscillating phase achieves the highest degree of mixing.} \\

\end{tabular}

 \end{@twocolumnfalse} \vspace{0.6cm}

  ]
%%%END OF TITLE, AUTHORS AND ABSTRACT%%%

%%%FONT SETUP - please do not change any commands within this section
\renewcommand*\rmdefault{bch}\normalfont\upshape
\rmfamily
\section*{}
\vspace{-1cm}

%%%FOOTNOTES%%%

\footnotetext{\textit{$^{a}$~Niels Bohr International Academy, Niels Bohr Institute, University of Copenhagen, Blegdamsvej 17, 2100 Copenhagen, Denmark}}

%Please use \dag to cite the ESI in the main text of the article.
%If you article does not have ESI please remove the the \dag symbol from the title and the footnotetext below.
%\footnotetext{\dag~Supplementary Information available: [details of any supplementary information available should be included here]. See DOI: 10.1039/cXsm00000x/}
%additional addresses can be cited as above using the lower-case letters, c, d, e... If all authors are from the same address, no letter is required

\footnotetext{\ddag~jonas.berx@nbi.ku.dk}

%%%MAIN TEXT%%%%

\section{Introduction}\label{sec:intro}
Mixing plays a fundamental role in a wide range of scientific and technological applications~\cite{Nguyen2005,HESSEL2005,Villermaux2019}, from large-scale industrial processes to microfluidics present in biological or medical systems~\cite{El-Ali2006,Stott2010,Fakhri2014}. At macroscopic scales, turbulent flows can promote efficient mixing through the stretching and folding of material interfaces, enhancing homogenisation~\cite{Aref2017,Thiffeault2006}. However, on smaller scales, flows are often characterized by low Reynolds numbers, where inertial effects are negligible, and mixing is dominated by diffusion and laminar advection~\cite{Ward2015}. Since the time required to mix over a distance $L$ is given by $t_{\rm mix} \sim L^2/D$ with $D$ the diffusion coefficient, diffusive mixing is inefficient for typical system on those length scales. This presents a significant challenge and necessitates the development of alternative strategies to enhance transport and homogenisation in such systems.

One promising approach to overcoming these limitations is the use of active particles to stir fluids~\cite{Jalali2015,Aref2017}. Active matter, consisting of self-propelled agents that generate motion at microscopic scales, can create flow fields that enhance mixing even in low Reynolds number environments. Biological systems, such as swimming microorganisms~\cite{Berg2004,Sommer2017,Visser2007} or cytoskeletal filaments driven by motor proteins~\cite{Stein2021}, naturally exploit active motion to facilitate transport and mixing at small scales~\cite{Shelley2024}. Inspired by these natural systems, synthetic active matter could potentially provide a means to drive efficient mixing in microfluidic applications~\cite{GROSJEAN201884}. Unlike externally driven flows, which rely on pumps or pre-defined channel geometries, active matter can generate self-sustained flows that dynamically adapt to the environment. Moreover, the inherent collective dynamics of active particles can lead to emergent coherent structures~\cite{Haller20215,Si2024}, which influence mixing in ways that differ significantly from traditional passive advection mechanisms.

To evaluate mixing efficiency various metrics have been developed. Classical measures include the variance of a passive scalar concentration field, or the finite-time Lyapunov exponent (FTLE), which quantifies stretching and separation of initially close trajectories~\cite{Tan2019}. More sophisticated measures such as multiscale norms, including the mix-norm or negative Sobolev norms, have been introduced to capture mixing dynamics beyond classical variance-based approaches~\cite{Mathew2003,Thiffeault2012}. These norms decay even in the absence of diffusion, making them suitable for assessing mixing in purely advective settings. However, all of these approaches typically require access to either density or velocity field information, which may not always be available, particularly in particle-based models, or in the case where only sparse Lagrangian data is available through particle trajectories~\cite{Thiffeault2010_2}. Since active matter systems often lack a well-defined velocity field due to the discrete nature of the particles and their interactions, alternative approaches are necessary to characterize their mixing properties effectively.

To address this limitation, an alternative approach is to quantify mixing using braid-topological methods based solely on particle trajectories. One such measure is the finite-time braiding exponent (FTBE)~\cite{Budišić2015}, which characterizes the complexity of particle exchanges in a flow by constructing braids from trajectories, and which provides a lower bound on the topological entropy, i.e., the mixing efficiency of the system. Moreover, the FTBE provides a measure of mixing without requiring knowledge of the velocity field, making it particularly suitable for discrete particle systems. Another related quantity is the writhe, which captures the geometric twisting of particle trajectories and provides insight into the underlying rotational transport mechanisms. 

In oceanography, for instance, such topological methods have been employed to analyse float trajectories and infer large-scale mixing properties of oceanic currents~\cite{Thiffeault2010_2}. Similarly, braid-based analyses of sparse particle trajectories have been used to characterize blood flow in the heart, revealing that healthy ventricular flow exhibits highly effective topological mixing, while deviations from this pattern correlate with reduced energetic efficiency in diseased states~\cite{Dilabbio2022}.

In this paper, we study the mixing properties of the chiral Vicsek model under confinement~\cite{Vicsek1995,Ginelli2016}, a paradigmatic active matter system, by analysing particle trajectories through the lens of braiding theory. The chiral Vicsek model consists of self-propelled particles that align with their neighbours and the boundary while undergoing constant-speed circular motion~\cite{Liebchen2022}. The interplay of collective alignment and chirality gives rise to complex emergent patterns~\cite{Lei2023,Caprini2024,Levis2018,Beppu2021} with different mixing properties, which are of seminal importance for the optimal design of future applications in, e.g., microfluidics or synthetic active matter. By constructing braids from particle trajectories, we compute the FTBE and writhe to assess the efficiency and characteristics of mixing in this system, and subsequently optimise the model parameters to deduce the emergent pattern that yields the highest degree of mixing efficiency.

The set-up of this paper is as follows. In Section~\ref{sec:model} we introduce the chiral Vicsek model and our braiding formalism, showing the different steady-state patterns and illustrating how to map spacetime trajectories into algebraic braids. In Section~\ref{sec:mixing}, the braid-topological measures relevant for mixing are introduced an subsequently computed for the different observed patterns. In Section~\ref{sec:optimisation}, we show a Pareto-optimal front between the writhe and FTBE before individually optimising the FTBE and numerically computing the parameters associated with the optimal mixing pattern. {\color{black}In Section~\ref{sec:roadmap}, we compare some of our computed patterns to experimentally observed ones ---in both artificial and living systems--- and propose a possible experimental system that could be used as a testbed to demonstrate the validity of our results.} Finally, in Section~\ref{sec:conclusions}, we conclude and provide avenues for future research.

\section{From active motion to braids}\label{sec:model}

\subsection{The chiral Vicsek model}

We consider a system of $N$ identical, self-propelled particles moving with a fixed constant speed $v_0$ inside a circular domain with radius $R$, similar to the model used in Ref.~\cite{Negi2023}. The particles interact through a polar interaction with strength $\gamma_p$, aligning its own orientation according to the average orientation of nearest neighbours within a circle of radius $\epsilon$. 

\begin{figure}[tp]
    \centering
    \includegraphics[width=\linewidth]{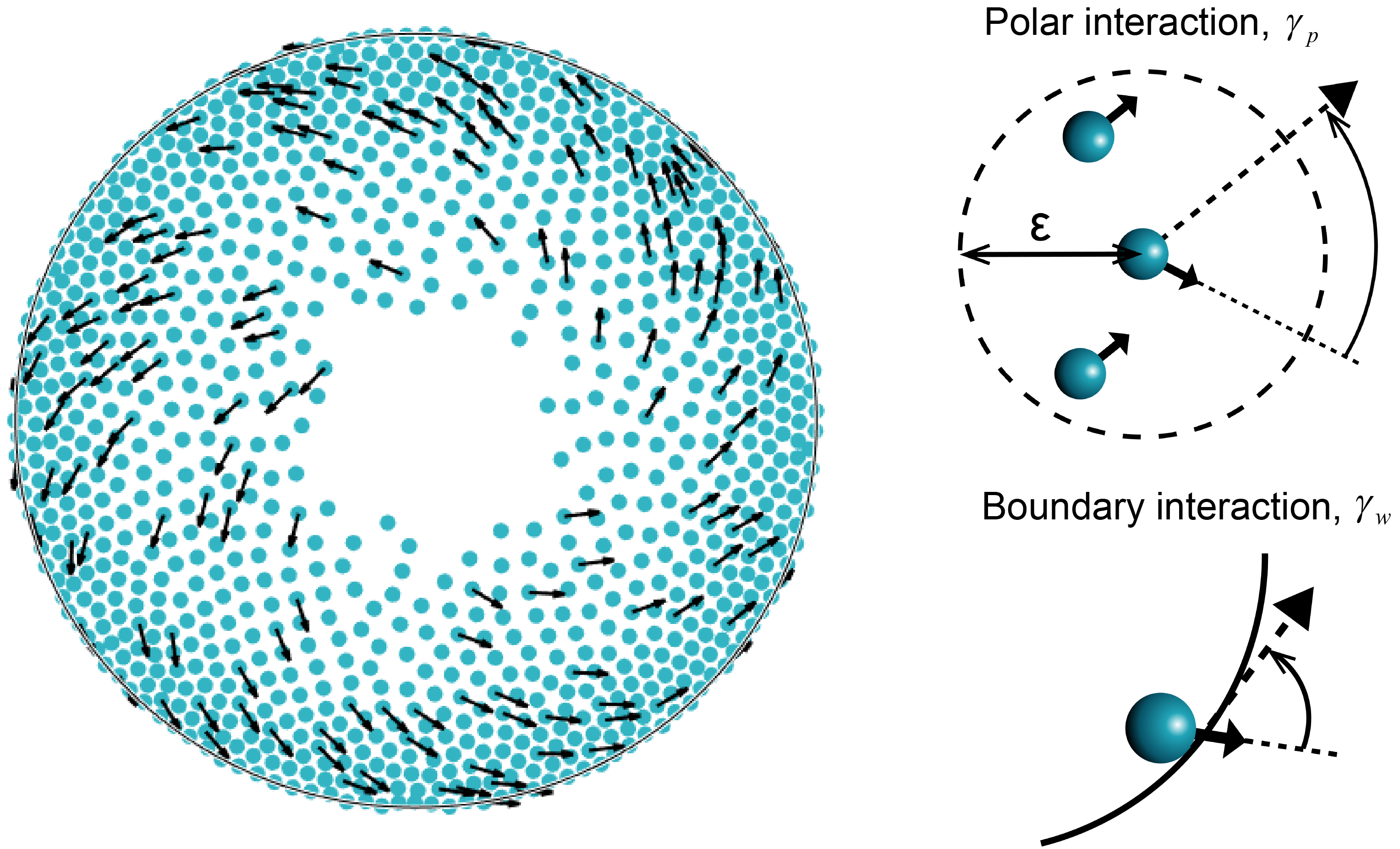}
    \caption{Snapshot of the chiral Vicsek model confined to a circular boundary, with the local orientation of a subset of active particles shown by arrows. The polar interaction (top right) between particles within a radius $\varepsilon$ re-orients the movement from an initial (dotted line) to a new (dashed line) direction. Particles colliding with the boundary (bottom right) re-orient their initial movement direction (dotted line) to one parallel to the boundary tangent (dashed line). Simulation parameters are $N=1000$, $\omega=0.45$, $\gamma_p = 0.5$ and $\gamma_w = 10$; the rest of the parameters is listed in the main text.}
    \label{fig:model}
\end{figure}

After a collision with the boundary, the particles align to it with strength $\gamma_w$, and there is a soft repulsive interaction both between particles with length scale $\ell$, and between particles and the wall with strengths $\kappa$ and $\kappa_w$, respectively. Both alignment interactions are shown in Fig.~\ref{fig:model}. The individual particle chirality is given by $\omega$, which encodes the inherent tendency to rotate in one direction, breaking rotational symmetry.

Given these interactions, the coupled equations of motion {\color{black}(EOM)} for the particle positions $\mathbf{r_m}(t) = (x_m(t),\,y_m(t))$ and orientations ${\theta_m}$, with $m \in\{1,\dots,N\}$, are given by

\begin{equation}
    \label{eq:EOM}
    \begin{split}
    {\bf \dot{r}_m} &= {\bf v}(\theta_m) + \frac{2\kappa}{\ell}\sum\limits_{r_{mn}<\varepsilon}({\bf r_m} - {\bf r_n})\,\mathrm{e}^{-\left(r_{mn}/\ell\right)^2}- \kappa_b \hat{\bf r}_m\, H(r_m-R)\,,\\
    \dot{\theta}_m &= \omega -\gamma_p \sum\limits_{r_{mn}<\varepsilon} \sin{(\theta_m-\theta_n)}-\gamma_w \sin{2(\theta_m-\phi_m-\frac{\pi}{2})}H(r_m - R) + \eta_{m}\,,
\end{split}
\end{equation}
with ${\bf v}(\theta_m) = v_0 (\cos\theta_m,\sin\theta_m)$ and where $H(x)$ is the Heaviside step function and $\eta_m$ is Gaussian noise with mean zero and two-point correlation $\langle \eta_m(t) \eta_n(t')\rangle = 2 D \delta_{mn}\delta(t-t')$. For all subsequent simulations we fix $v_0 = 1$, $R =12$, $\kappa = 3$, $\kappa_b = 20$, $\ell = 0.3$, $\varepsilon = 1$ and $D = 0.01$. This particular choice of parameters aligns well with experimental data on \emph{Escherichia coli} suspensions confined within a microwell~\cite{Beppu2021}. The remaining tunable parameters in our system are thus $N,\,\omega,\,\gamma_p$ and $\gamma_w$.

\begin{figure*}[htp]
    \centering
    \includegraphics[width=\linewidth]{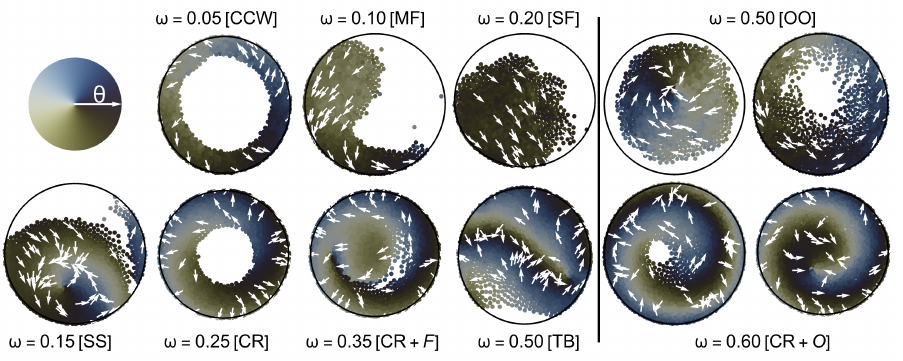}
    \caption{The different representative NESS patterns observed in this work. The top row displays patterns in the weak alignment regime for fixed $N = 1000$ and increasing $\omega$, while the bottom row illustrates the corresponding patterns in the strong alignment regime for fixed $N = 2000$. A colour legend indicating the orientational angle of the active particles is shown in the upper left corner. Abbreviations for each steady-state pattern are given in square brackets, with their full descriptions provided in the main text. For both rows, patterns on the right of the vertical line (OO and CR+O) oscillate periodically between the two side-by-side patterns.}
    \label{fig:steady_state}
\end{figure*}

We initialise the system by choosing the orientations of all particles isotropically, i.e., parallel to the radial direction, so there is no inherent bias in the rotational direction, aside from the particles' chirality. The Euler-Mayurama iteration scheme is then used to solve the EOM. Simulations are run with a timestep of $\Delta t = 0.01$ for a total time of at least $t = 1500$, or until a steady state has been reached. For every choice of parameter combinations, 10 independent simulations with random initial conditions are performed. For every simulation, we record Lagrangian data on 500 tagged particles, together with their orientations, at sampling intervals $\Delta t_s = 1$ for a total time of $T = 500$ in the steady state.

A brief remark on the chosen boundary conditions is in order. Two ubiquitous choices are periodic or fixed boundary conditions. While the former is useful for simulating infinite systems, the latter pertains more to biological or medical settings, where e.g., bacteria are confined to move within spatial boundaries such as cells or larger tissues, or for the mixing of chemicals within a microfluidic device. The influence of periodic boundary conditions on the patterns found in the chiral Vicsek model was discussed in Ref.~\cite{Beppu2021}; they generally lead to the non-existence of some patterns. For instance, spiral droplets are stabilised under the influence of confinement, while only being transient patterns for infinite systems. Furthermore, periodic boundaries fundamentally change the braid description of the trajectories, since particles can now make a full tour around the vertical or horizontal directions, which introduces additional braid operations~\cite{Finn2007}; we will not consider such boundary conditions in this work.

\subsection{Steady-state patterns}

The interplay between the number of particles $N$, the chirality $\omega$ and the alignment interactions $\gamma_p,\,\gamma_w$ leads to a rich variety of non-equilibrium steady-state (NESS) pattern formation, as explored in detail through numerical simulations in Ref.~\cite{Negi2023}. While a complete characterization of the steady-state behaviour is beyond our scope, we briefly comment on the typical NESS patterns observed. In the sections that follow, we will distinguish between weak and strong alignment regimes, defined respectively by $\gamma_p=0.1,\, \gamma_w =1$ and $\gamma_p=0.5,\, \gamma_w =10$. Although this binary classification does not capture the full diversity of possible patterns, it serves as a representative framework for the most commonly encountered cases, and we adopt it throughout to organize our discussion of steady-state behaviour. In Sec.~\ref{sec:optimisation}, however, we perform a broader exploration of parameter space through a heuristic optimization of the mixing behaviour. 

In the weak alignment regime and for moderate values of 
$N$ (see Fig.~\ref{fig:steady_state} top row), we observe that as $\omega$ increases, the NESS patterns undergo a sequence of transitions: from counterclockwise edge currents (CCW)—where particles circulate along the boundary—to flocking states, and eventually to a state characterized by ordered oscillations (OO) between a CCW edge current and a central aggregate where particles collectively rotate in the clockwise (CW) direction~\cite{Caprini2024}. The flocking state can be divided further into one wherein particles form multiple flocks (MF), approximately at $\omega = v_0/R = 1/12$, and another in which a single flock (SF) emerges after a transient spiral flocking state. For higher $N$, the OO state disappears. 

In the strong alignment regime (see Fig.~\ref{fig:steady_state} bottom row), both CW and CCW edge currents reappear at low values of $\omega$ across all $N$ (not shown). As $\omega$ increases, the system transitions either to a flocked state—MF followed by SF for low $N$, while predominantly SF for moderate $N$—or to a state where a stable spiral (SS) droplet persists in the NESS, for high $N$. For low $N$, increasing $\omega$ reproduces behaviour similar to the weak alignment regime and the systems transitions into the OO phase. For moderate to high $N$, however, the NESS patterns transition from either the SF or SS into a state in which counterrotating currents (CR) emerge; due to the strong boundary alignment dominating over the chirality, CW edge currents emerge, with particles further away from the boundary forming a CCW rotating current. A single-vortex state with a counterrotating boundary layer have been observed in, e.g., confined bacterial suspensions~\cite{Wioland2013}. Increasing $\omega$ further leads the particles away from the boundary to form flocks, resulting in a combination of counterrotation and flocking (CR+F), and eventually for very high $\omega$ this leads to counterrotation with oscillation (CR+O), similar to the OO phase, in which a stable spiral is periodically formed and then destroyed near the centre. For moderate $N$, however, this final phase is observed only for very high $\omega \gtrsim 0.6$; for $\omega \lesssim 0.6$ particle oscillations are unstable and the system collapses into travelling bands (TB) that rotate around the system instead.

\subsection{Braiding trajectories}
\begin{figure}[htp]
    \centering
    \begin{subfigure}{0.55\linewidth}
        \centering
        \includegraphics[width=\linewidth]{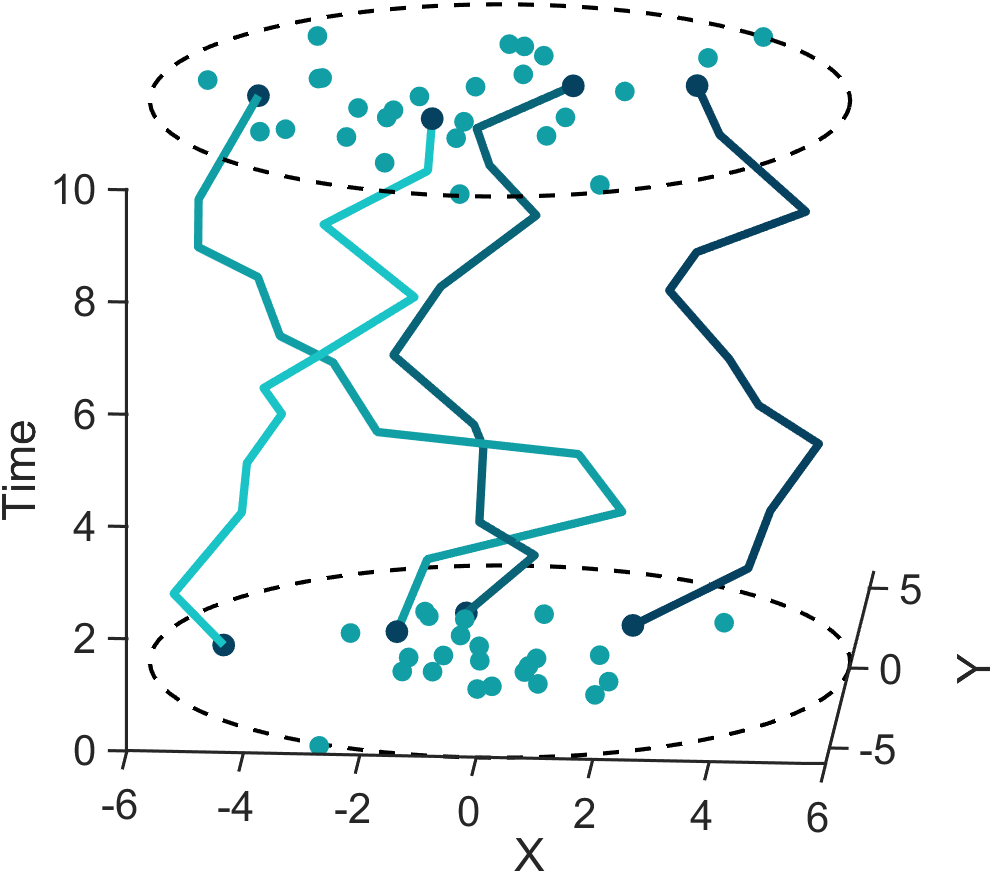}
    \end{subfigure}
    \begin{subfigure}{0.44\linewidth}
        \centering
        \includegraphics[width=\linewidth]{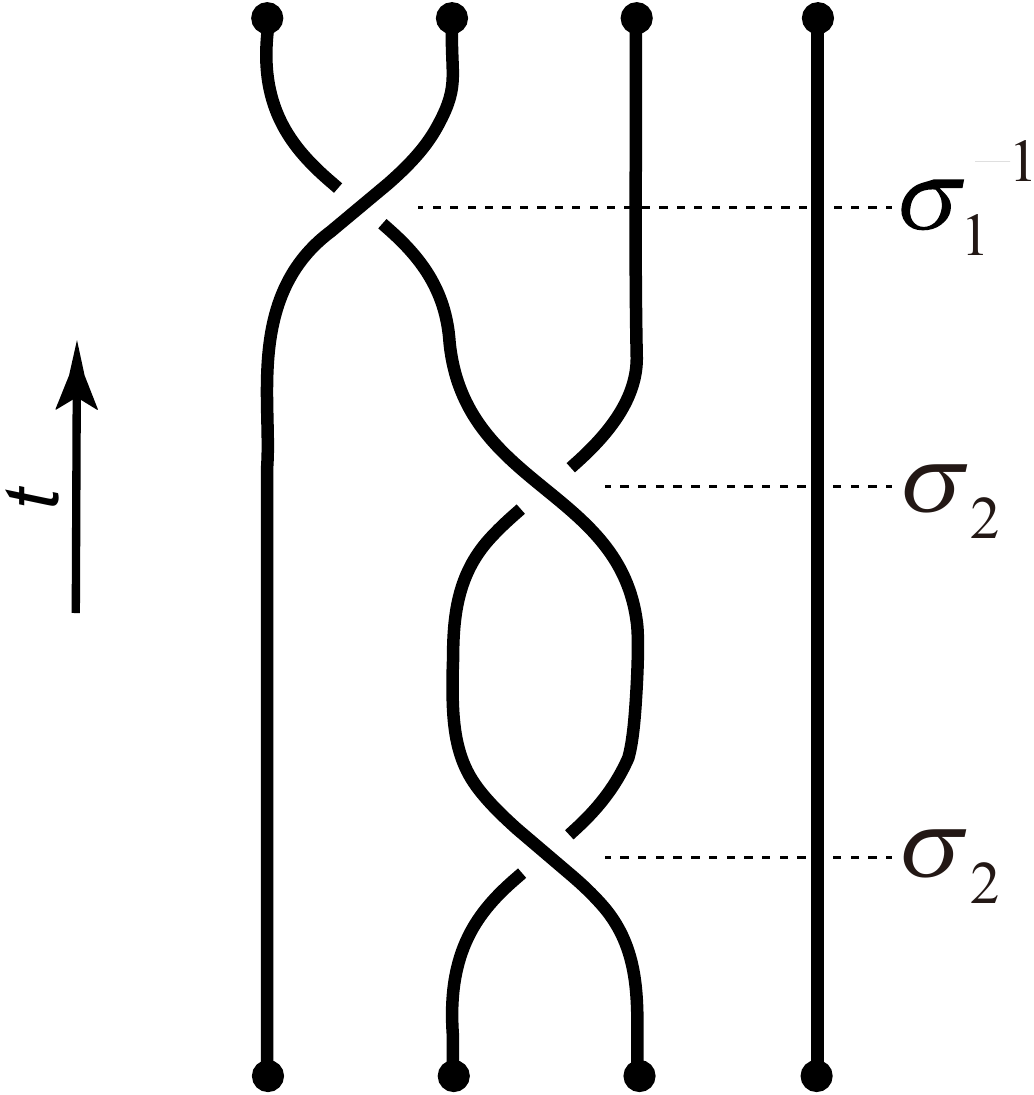}
    \end{subfigure}
    \caption{Mapping of spacetime trajectories of four tagged particles (\emph{left}) to an algebraic braid (\emph{right}) by projecting the trajectories on the $x-t$ plane. The resulting braid is $\beta_4 = \sigma_2^2 \sigma_1^{-1}$ with writhe $\rm{Wr} = 1$ and braid length $L = 3$.}
    \label{fig:trajectories}
\end{figure}

To characterise mixing, we record the spacetime trajectories of a number $n\leq N$ tagged particles and project the trajectories onto a two-dimensional subspace, where one of the dimensions is the temporal coordinate~\cite{Thiffeault2010_2}. Henceforth, let us assume we project onto the $x-t$ plane. When projecting, particle trajectories can cross at different points in time. Although we projected on the $x-t$ plane, the remaining spatial coordinate $y$ determines whether a single trajectory crosses either behind or in front of others. In this manner the trajectories become entangled, and the entanglement information is encoded in the collection of crossings between particle trajectories. Based on the temporal coordinate, these trajectory crossings can then be sorted into a standard form, where crossings cannot occur at the same time, neglecting the length between them. {\color{black} Note that a single crossing always involves two \emph{adjacent} strands in the standard braid form; entangling distant strands necessarily produces additional crossings with intermediate strands due to the projection operation. After each crossing, the strands are relabelled such that the leftmost strand has once again index $1$.}
The resulting object is an \emph{algebraic braid}, which contains all topological information of the particle movement. A visual representation of this process for our model is shown in Fig.~\ref{fig:trajectories}, where four particles entangle to form a braid.

A crossing of strands $i$ and $i+1$ can be represented by operators $\sigma_i^{\pm 1}$, where the positive (negative) exponent indicates {\color{black}that strand $i$, counted from the left, crosses under (over) strand $i+1$}. These operators can be concatenated into a representation of the braid: $\beta_n = \sigma_i^{\pm\nu_i} \sigma_j^{\pm\nu_j}...\sigma_k^{\pm\nu_k}$, where all $\nu = \pm 1$. To convert the trajectories into braids we use the \texttt{braidlab} Matlab package~\cite{Thiffeault2022}, which also allows for the calculation of a myriad of braid measures.

To collect sufficient statistics over a complete range of braid sizes $n$ for a single parameter choice, we first construct 10 large individual braids from the 10 independent simulations including all 500 tracked particles. Subsequently, we extract 100 subbraids of size $n$ per braid by randomly selecting $n$ strands in the braid and removing the rest, for a total of 1000 braids of size $n$ for every parameter choice. This subbraid ``bootstrapping'' approach to braid statistics has been shown to be identical to performing individual simulations with $n$ tagged particles~\cite{Budišić2015}, greatly decreasing the computational load. 

\section{Braided mixing}\label{sec:mixing}

\subsection{The braid writhe}
From the braid word $\beta_n$, we can compute the writhe $\rm{Wr} = \sum_i \nu_i$, which is simply the sum of the exponents $\nu$ in the braid representation. The writhe is a measure for the global twist of the braid or, in our system, for the global rotation of the system within the observed time window; the sign indicates the overall direction of rotation and the amplitude is a measure for the rotation strength. Due to our choice of projection plane, CW entangled particle trajectories correspond to a positive writhe, while CCW entanglement yields a negative writhe. If only two particles are selected and a braid is constructed from their trajectories, the writhe equals twice the number of times one particle circles the other. Since the writhe is a topological quantity, it counts the \emph{number} of times particle trajectories wind within a particular time window, independent of the distance travelled. If each particle interacts with every other particle in a single period of rotation, the length of the resulting braid would scale as $L \sim n (n-1)/2$. Since for $\omega >0$ a global CCW rotation is induced, the writhe will scale roughly with the length of the braid. As such, a rescaled braid writhe, $\rm{Wr}_n = \rm{Wr}/n^2$, provides good data collapse, and we henceforth only use this this quantity.

\begin{figure}[htp]
    \centering
    \includegraphics[width=\linewidth]{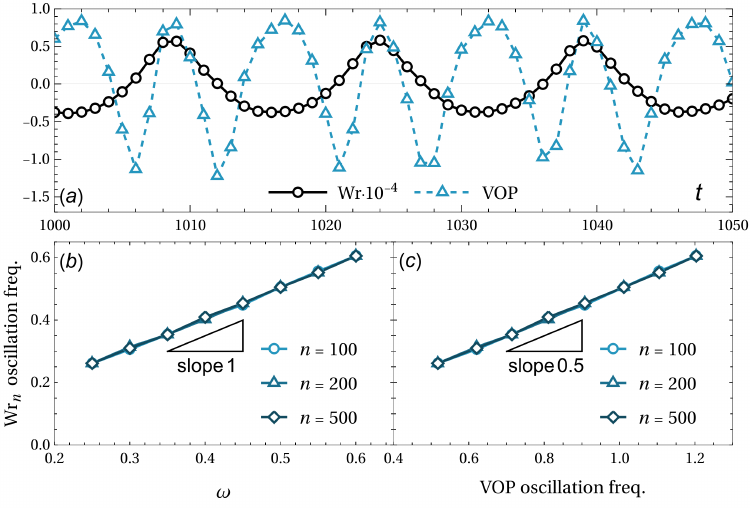}
    \caption{{\bf{(a)}} The periodic motion of the braid writhe (black) and VOP (blue) as a function of time in the high alignment regime ordered oscillation regime for $N=1000$ and $\omega =0.4$. It can be seen that the period of oscillation of the writhe is almost exactly twice as long as the VOP's. The writhe $\rm{Wr}_n$ oscillation frequency is shown as a function of the chirality {\bf (b)} and the VOP oscillation frequency {\bf (c)} for $N=500$, with the slope of the linear relation indicated.}
    \label{fig:oscillation_Wr_VOP_H}
\end{figure}

We will compare the braid writhe to an order parameter used for the classification of circular phases and patterns: the vortex order parameter (VOP), given by
\begin{equation}
    \mathrm{VOP}=\frac{\langle\cos \phi\rangle-2 / \pi}{1-2 / \pi} = \left(\frac{1}{N}\sum_{i=1}^N\frac{\left|\mathbf v_i \cdot \mathbf{T}_i\right|}{\left\|\mathbf v_i\right\|}-\frac{2}{\pi}\right)\frac{1}{1-2 / \pi}\,,
    \label{VOP}
\end{equation}
where the angle $\phi$ is between the particle velocity $\mathbf v_i$ and the unit {\color{black}vector $\mathbf{T}_i$ in the azimuthal direction}, at the position of the particle $i$. The VOP is equal to one when the system exhibits perfect vortex dynamics and equal to zero when there is perfect disorder. It becomes negative when the motion is radial. Edge currents, for instance, have $\rm{VOP} \approx 1$, while oscillatory patterns transition between circular motion $(\rm{VOP} \approx 1)$ and radial motion $(\rm{VOP} < 0)$.

In Fig.~\ref{fig:oscillation_Wr_VOP_H}, we show that the braid writhe and VOP are directly related to the chirality. The oscillation frequency of the writhe is exactly equal to $\omega$, while the VOP frequency is $2\omega$. This relation illustrates the usefulness of the writhe over the VOP; whereas the latter requires full knowledge of the velocity field to compute the instantaneous orientation and rotation of fluid elements for the VOP, the braid writhe can be obtained directly from the topology of discrete particle trajectories alone, making it particularly well-suited to systems where only sparse Lagrangian data are available.

\subsection{The finite-time braiding exponent}
To characterise topological complexity or mixing, the writhe is not a good measure. To see this, one can simply consider the previous example of two particles circling each other. Longer trajectories lead to a linearly increasing writhe without increasing the topological complexity, since both trajectories can still be trivially disentangled by rotating the embedding space. A similar argument can be made for the braid word length $L$, i.e., the number of operators $\sigma_i^{\pm 1}$ in a braid $\beta_n$. Simply inserting the combination $\sigma_i^+ \sigma_i^-$ for any $1\leq i<n$ increases the braid word length to $L+2$ without increasing the complexity since both operators can be eliminated to yield the unit operator.

\begin{figure}[tp]
    \centering
    \begin{subfigure}{0.45\linewidth}
        \centering
        \includegraphics[width=\linewidth]{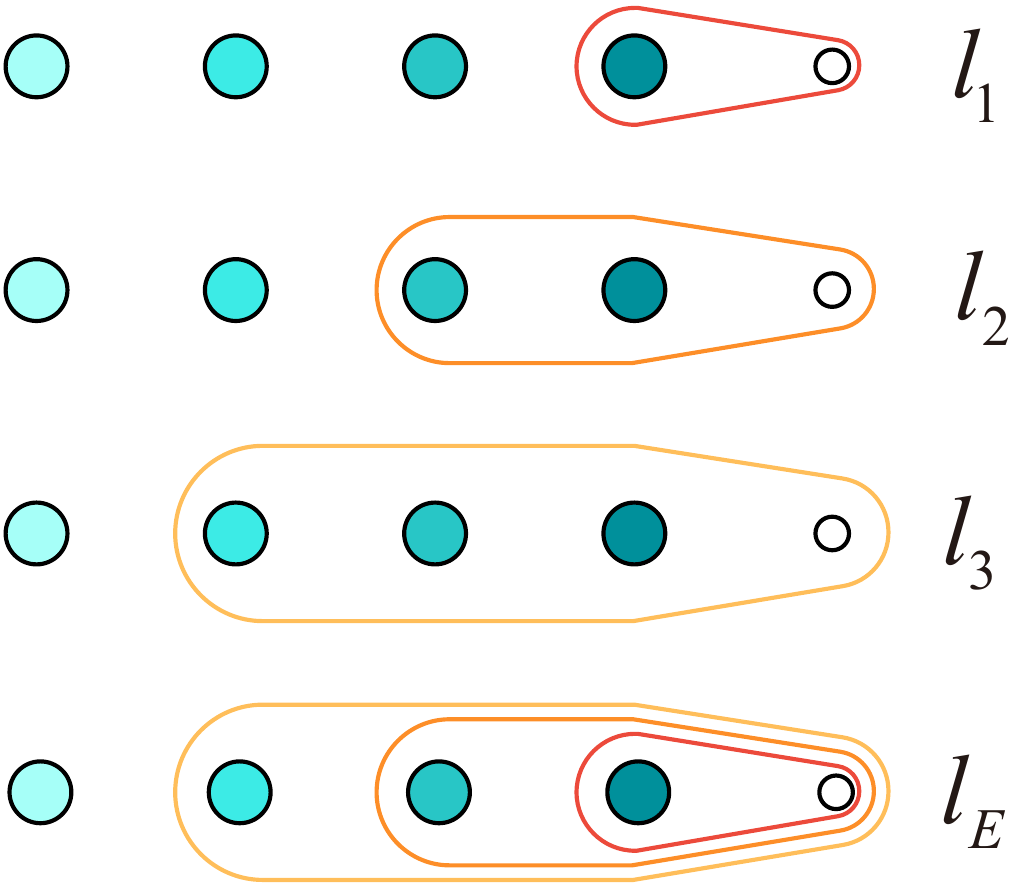}
    \end{subfigure}
    \rulesep
    \begin{subfigure}{0.49\linewidth}
        \centering
        \includegraphics[width=\linewidth]{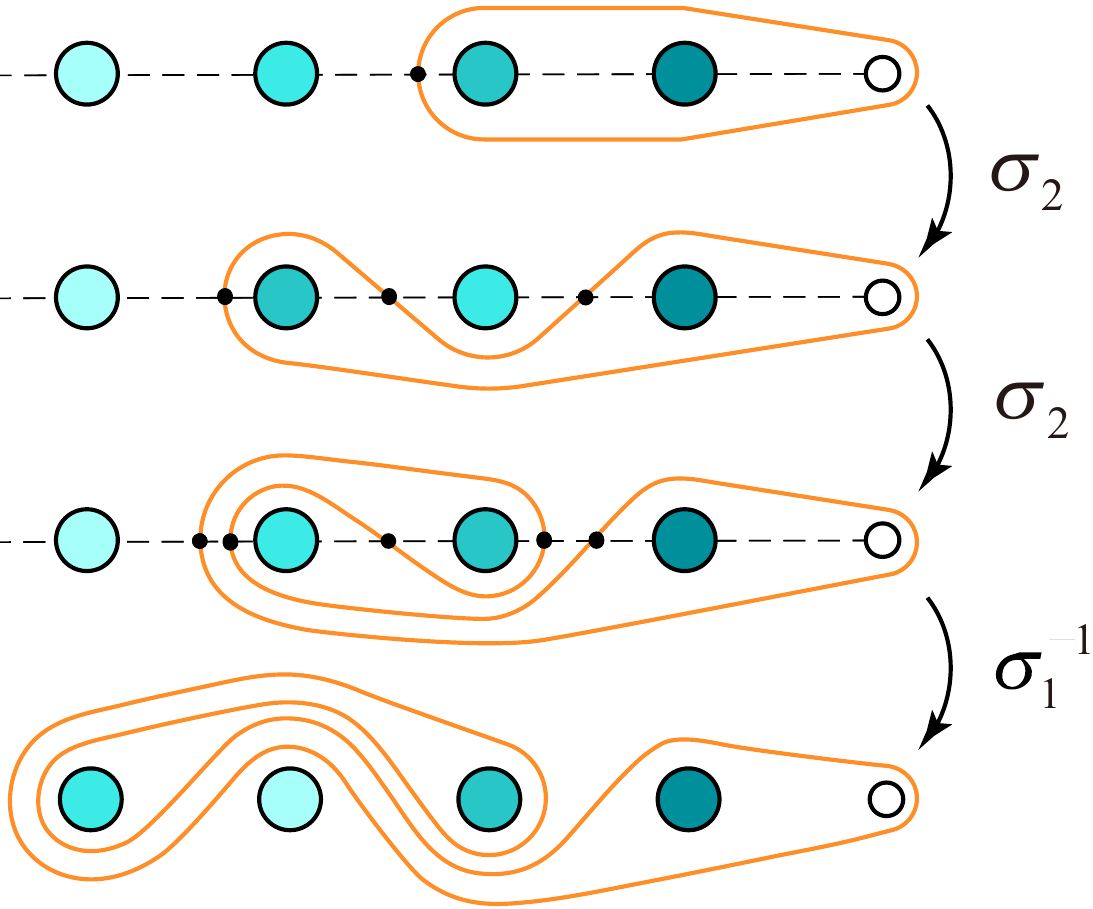}
    \end{subfigure}
    \caption{The definition of $\ell_E$ for a braid of four strands as loops around punctures (coloured circles) that cannot be deformed into a single point. {\color{black}The rightmost uncoloured point is the base point which does not participate in the braid formed and serves as an anchor for all loops.} The right panel illustrates the deformation of the loop $\ell_2$ (yellow) under the action of the braid $\beta_4 = \sigma_2^2 \sigma_1^{-1}$, for which representative trajectories are shown in Fig.~\ref{fig:trajectories}. {\color{black} The growth of the number of intersections (black points) between the deformed loop and the real axis (thin dashed line) is a measure of the growth rate of the material loop.}}
    \label{fig:loops}
\end{figure}

A more natural way to characterise mixing is to consider material loops surrounding the strands and studying the (exponential) stretching of the loops by applying the braid operators. The growth rate of a loop under the repeated action of a braid then gives the braid entropy. Let us consider a material loop $\ell_E$ representing a generating set of the non-oriented fundamental group on the $n$-punctured disk~\cite{Moussafir2006}, i.e., a loop around ``holes'' in a disk representing the braid strand starting points, see the left panel of Fig.~\ref{fig:loops}. {\color{black}The loops also encircle an additional point, represented as an empty circle in Fig.~\ref{fig:loops}, which does not participate in the braiding but serves as an anchor for the loops to wrap around.} Acting on the punctures with a braid operator $\sigma_i^{\pm 1}$, rotating them in the plane, can lead to the deformation and subsequent stretching of the loop, see the right panel of Fig.~\ref{fig:loops} for the braid $\beta_4 = \sigma_2^2 \sigma_1^{-1}$. The resulting integral laminations (the set of disjoint non-homotopic simple closed curves) formed by the action of the braid can be encoded by using Dynnikov coordinates~\cite{Dynnikov2002}. 

{\color{black}The extent to which a sequence of operators folds the material loops can be measured by counting the number of intersections of the loops with the horizontal real axis through the punctures, and tracking how this number changes after applying each operator. This is shown in the right panel of Fig.~\ref{fig:loops} as black points, the number of which grows after each application of a braid operator. While the figure illustrates this for a single loop $\ell_2$, the total number of intersections for all loops $\ell_E$ can be counted by the norm $|.|$. In particular, for a braid $\beta_n$ consisting of a sequence of operators, the number of intersections of all loops with the real axis is given by $|\beta_n \ell_E|$. The relative growth of this number is then a measure for the growth rate of the material lines, and hence the mixing.}

To characterise this growth, the FTBE of a braid $\beta_n$ of $n$ particles for an interval $T$ is defined as~\cite{Budišić2015} 
\begin{equation}
    {\rm FTBE}_n = \frac{1}{T} \log\frac{|\beta_n\ell_E|}{|\ell_E|}
\end{equation}
and measures the inverse time to entangle $n$ particles through the growth rate of the material lines. The FTBE serves as a proxy and lower bound for the topological entropy of planar dynamical systems. When the number of strands in the braid increases, the FTBE approximates the topological entropy better and effectively describes the degree of mixing on different scales; by computing a \emph{spectrum} of FTBE's for different $n$, we can in principle compare local and global mixing efficiency~\cite{Thiffeault2010_2,Budišić2015}. 

\begin{figure}[htp]
    \centering
    \includegraphics[width=\linewidth]{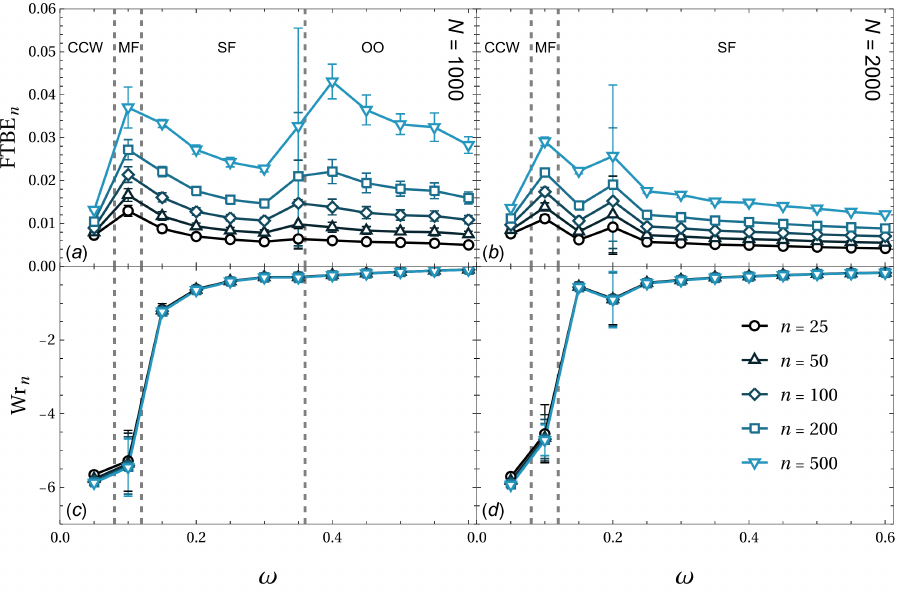}
    \caption{The FTBE$_n$ (a,b) and rescaled writhe (c,d) as a function of chirality for $N=1000$ (a,c) and $N=2000$ (b,d) in the weak alignment regime where $\gamma_p = 0.1,\, \gamma_w = 1$. The rescaled writhe provides a very good data collapse. Regions of chirality where certain patterns occur are approximately delineated by the vertical dashed lines using Ref.~\cite{Beppu2021} and visual inspection. The patterns are abbreviated as follows: counterclockwise edge current (CCW), multiple flocks (MF), single flock (SF) and orderded oscillation (OO).}
    \label{fig:wr_ftbe_low}
\end{figure}

In the weak alignment regime, where $\gamma_p = 0.1,\,\gamma_w = 1$, the rescaled writhe and FTBE are shown in Fig.~\ref{fig:wr_ftbe_low}. It can be seen that in the counter-clockwise edge current phase (CCW), the writhe is strongly negative, indicating that the system moves as a whole in the CCW direction, while the FTBE is low for all $n$. This shows that uniform stirring, i.e., simple rotation, does not induce mixing in the system; material lines simply stretch with the stirring rods but do not fold, since the relative positioning of the active particles is preserved in the boundary vortex. 

Increasing the chirality to $\omega \gtrsim v_0/R$, the system forms multiple flocks (MF), which still rotate collectively in the CCW direction. Once again, the writhe is strongly negative, but in this regime the mixing efficiency increases significantly; flocks can merge, separate or perform complex rotations around each other which braids the trajectories in a complex manner, even though within a single flock the relative position of particles with respect to each other is preserved. A slight increase of $\omega$ induces a collective state where the flocks aggregate and form a single large flock (SF); particles only locally rearrange within the flock but keep a global orientation. The flock moves slowly along the system boundary in the CCW direction, which present a state with small but finite writhe, and a decreasing FTBE. 

Increasing the chirality lowers the absolute value of the writhe, such that $|{\rm Wr}_n |\downarrow 0$, as well as the FTBE, see Fig.~\ref{fig:wr_ftbe_low}(b,d). Due to a decreasing rotation radius for individual particles when $\omega$ increases the system jams and the single cluster is confined to the system boundary without moving. However, when the particle fraction is low enough that particles can complete their individual circular motion, a jammed state is avoided, however, and the system exhibits collective ordered oscillations (OO). Since this collective motion is rotationally symmetric the writhe is zero on average, see Fig.~\ref{fig:wr_ftbe_low}(c). However, due to the complex interaction between local rotation and collective motion, the mixing is significantly increased, resulting in a sharp increase in the FTBE (Fig.~\ref{fig:wr_ftbe_low}(a)).

\begin{figure*}[htp]
    \centering
    \includegraphics[width=\linewidth]{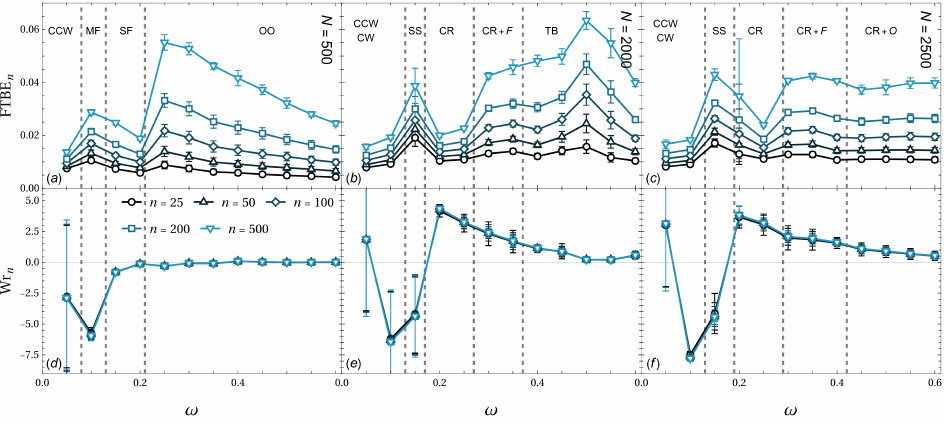}
    \caption{The FTBE$_n$ (top) and rescaled writhe (bottom) as a function of chirality for $N=500, \,2000, \, 2500$ in the strong alignment regime where $\gamma_p = 0.5,\, \gamma_w = 10$. Regions of chirality where certain patterns occur are approximately delineated by the vertical dashed lines using Ref.~\cite{Beppu2021} and visual inspection. The patterns are abbreviated as follows: (counter)clockwise edge current (CCW/CW), multiple flocks (MF), single flock (SF), ordered oscillation (OO), stable spiral (SS), traveling bands (TB), counterrotation (CR) plus flocking (+F) or plus oscillations (+O).}
    \label{fig:wr_ftbe_high}
\end{figure*}

In the strong alignment regime, with $\gamma_p = 0.5, \,\gamma_w = 10$, other steady-state patterns emerge that exhibit more exotic mixing behaviour, as was shown in the bottom row of Fig.~\ref{fig:steady_state}. For very small values of $\omega$, the particle alignment with the wall dominates over the intrinsic chirality, such that for increasing particle fraction the mean writhe increases, due to a higher weight of clockwise (CW) edge currents in the ensemble; this leads to large fluctuations as illustrated in the bottom row of Fig.~\ref{fig:wr_ftbe_high}. For the corresponding FTBE, however, this is irrelevant and the same low-mixing behaviour as in the weak alignment regime holds. 

For low particle fractions, as in Fig.~\ref{fig:wr_ftbe_high}(a,d), the same behaviour as a function of $\omega$ holds, with different phase boundaries. For higher fractions however, the picture changes. For intermediate densities, the MF state is non-existent and the system directly transitions into the SF regime. 

For intermediate-to-high densities (Fig.~\ref{fig:wr_ftbe_high}(b,c,e,f)), the system instead transitions into a sustained spiral droplet (SS), which on average rotates in the CCW direction, i.e., with a negative mean writhe, and which moves along the system boundary. Particles on either side of the point of contact of the droplet with the boundary show opposite chiral motion, which sustains the droplet and ultimately leads to a higher degree of mixing due to complex braiding interactions within the droplet. 

Increasing the chirality leads to counterrotating currents (CR); particles far from the centre align in the CW direction due to the strong boundary interactions, while for particles closer to the centre the polar interactions lead to collective CCW currents, effectively separating the system into two subsystems with a soft boundary in between where the motion is purely radial. Particles can be exchanged between the two counterrotating regions to a minor extent, leading to a higher degree of trajectory entanglement and FTBE than simple edge currents, but significantly lower than other regimes. 

Increasing $\omega$ even more leads to a phase of counterrotation with flocking (CR+F). Due to the counterrotation, flocks can move in opposite directions or break apart due to shearing interactions near the soft boundary in between the two regions of opposite rotation. Similar to the MF regime, the individual flocks can therefore exhibit complex dynamics, braiding the particle trajectories.

Finally, for very high chirality the system can either exhibit travelling bands (TB) or counterrotating currents with oscillating behaviour (CR+O), depending on the particle density, see Fig.~\ref{fig:wr_ftbe_high}(b,c,e,f). The CR+O state continues the downward trend of the writhe and maintains approximately a constant FTBE, the counterrotating vortex diminishing due to the intrinsic chirality becoming more dominant with respect to the boundary alignment strength. Travelling bands, however, seem to significantly increase the FTBE, up to a value of $\omega \approx 0.5$. Since particles within the same bands are strongly aligned, mixing is suppressed at local scales. At band interfaces, however, a high shear-like motion leads to better stretching and folding of material lines and the FTBE increases significantly as a result. The system eventually transitions into the CR+O regime when the chirality is increased even more.

Based on the previous analysis, it becomes clear which steady-state patterns generally lead to the lowest and highest mixing efficiencies. Simple uniform steady-state rotation, e.g., edge or counterrotating currents, as well as globally ordered polar states, i.e, single flocks lead to inefficient mixing due to both local and global conservation of relative particle ordering. On a topological level, the fraction of topologically unprotected braids in the ensemble increases for such patterns; such braids can be ``disentangled'' by rotating the space during time evolution. By moving to a co-rotating frame the system can be brought to a dynamical state without net movement, effectively suppressing any mixing.

Conversely, high mixing is achieved when the system either exhibits patterns involving multiple flocks (MF, CR+F), spirals, travelling bands or oscillatory motion (OO, CR+O). Note that in general the patterns exhibiting this high degree of mixing concomitantly show a low writhe (in absolute value); oscillatory motion and travelling bands exhibit a writhe that is close to zero, while for spiral and multiple flocking phases the writhe is still higher. This leads us to the hypothesis that for optimal mixing the braid writhe and FTBE are inversely related; the writhe seems to be more indicative of the mean circulation of the particles and is related to the stirring motion in the system. A low writhe indicates that there is (almost) no preferred direction for the stirring while a high writhe fixes a mean stirring direction. 

\section{Mixing and Pareto-optimal trade-offs}\label{sec:optimisation}

Let us test the assertion made in the previous section by simultaneously maximising both the magnitude of the rescaled writhe $\rm{Wr}_n$ and the FTBE$_n$. Since both quantities are related through a common set of system parameters, independent optimisation is generally not possible, leading to trade-offs between writhe and FTBE. A trade-off is Pareto-optimal if further optimisation of one objective is necessarily detrimental to at least one other objective. While Pareto fronts are commonly used in engineering, they have recently been used to determine, e.g., cost-error trade-offs in biological discrimination and proofreading processes~\cite{Chiuchiu2023,Berx2024}, optimal phenotypical fitness landscapes~\cite{shoval2012}, and morphogenetic trade-offs between information, fitness and cost~\cite{Rue2009,Berx2025}.

From a scalarised perspective, finding the Pareto trade-off is analogous to minimising the following free-energy-like quantity
\begin{equation}
    \label{eq:free_energy}
    \Omega_n = -\lambda|{\rm Wr}_n| - (1-\lambda) {\rm FTBE}_n\,.
\end{equation}
Doing so results in a parametric front of absolute writhe versus FTBE~\cite{seoane2015}. A concave front would suggest a protocol --a change in the underlying system parameters by tuning $\lambda$-- that transitions smoothly from a state maximising the writhe to one maximising FTBE. A convex front, however, suggests that only the two endpoints, i.e., the points with either maximal writhe or FTBE are optimal solutions, given that solutions lying in between are metastable in the sense that they are \emph{local} minima of equation~\eqref{eq:free_energy}, while the endpoints are \emph{global} minima. By tuning $\lambda$, a sudden switch from one endpoint to the other can then be achieved, similar to the liquid-gas first-order phase transition in classical thermodynamics. In between, for a linear front, any Pareto-optimal solution becomes optimal at a critical value of the parameter $\lambda$, resembling a first-order phase transition at criticality. Such a linear front implies a proportional trade-off between writhe and FTBE.

\begin{figure}[htp]
    \centering
    \includegraphics[width=\linewidth]{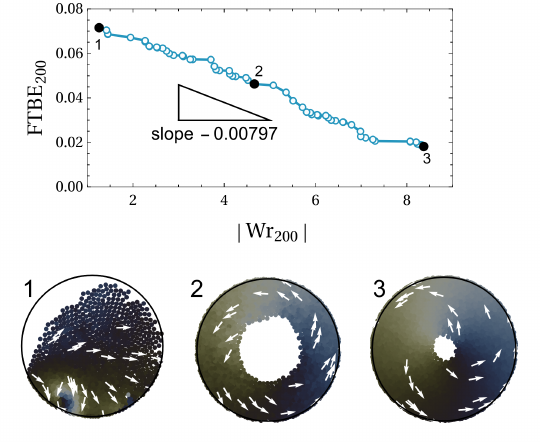}
    \caption{The Pareto-optimal trade-off (symbols) between the absolute value of the writhe and FTBE for $n=200$. The indicated slope is calculated by a linear fit with goodness-of-fit $R^2 = 0.987$. Patterns associated with the black labelled points are given underneath. Points 2 and 3 present CCW edge currents, while for point 1, which has the highest FTBE, the system evolves into an edge-confined vortex periodically colliding with a moving flock.}
    \label{fig:pareto}
\end{figure}

To compute the Pareto front, we use a NSGA-II elitist genetic algorithm~\cite{Deb2008}. Fig.~\ref{fig:pareto} shows that the trade-off between the writhe and FTBE for braids with $n=200$ strands is approximately linear, indicating that they are inversely related, confirming our initial suspicions. Increasing the writhe proportionally decreases the maximal mixing efficiency that can be achieved. Due to the computational load involved in numerically determining the Pareto front, however, the accuracy of the results is limited and possibly allows for other front shapes to emerge when accuracy can be improved. Hence, we will not definitively claim here that the front must be linear. {\color{black} The parameter ranges for the numerical optimisation are chosen as $\omega\in[0,0.6],\, N\in[500,3000],\,\gamma_p\in[0.1,...]$ and $\gamma_w\in[1,...]$; this choice fully covers both the weak and strong alignment regime for $\gamma_p,\,\gamma_w$, as well as the complete $\omega-N$ phase space determined in Ref.~\cite{Beppu2021}.} 

To check which steady-state patterns yield the highest mixing efficiency, however, we use the Nelder-Mead simplex method~\cite{Lagarias1998} to maximise the FTBE$_n$ independently of the writhe, {\color{black} with the same parameter ranges as in the Pareto optimisation}. The numerically determined optimal value is ${\rm FTBE}_{200} = 0.072$, which is very close to the leftmost endpoint labelled by number 1 in Fig.~\ref{fig:pareto}. The parameters corresponding to this state are listed in Table~\ref{tbl:parameters}, together with a second simulation for $n=50$ tracer particles, confirming the optimal parameter combination to a good degree. The concomitant optimal mixing pattern combines a stable vortex confined at the boundary with the rest of the particles exhibiting ordered oscillations. The system oscillates between the vortex acting as a mixer and subsequently particles being pushed out and new particles pushed in to be mixed. This interaction between the vortex and the rest of the system leads to the observed high degree of mixing.

\begin{figure}
    \centering
    \includegraphics[width=0.6\linewidth]{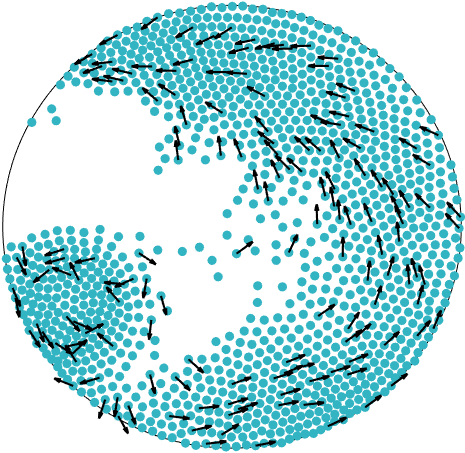}
    \caption{The emergent pattern associated with the maximisation of the FTBE with parameters given in Table~\ref{tbl:parameters}. A single vortex confined to the system boundary acts as a mixer; particles from the outside of the vortex are pushed inside in an oscillatory fashion, while particles in the vortex escape to the bulk.}
    \label{fig:optimal_mix_pattern}
\end{figure}

\begin{table}[]
\centering
\caption{Parameters corresponding to the optimal pattern yielding the highest mixing efficiency.}
\begin{tabular}{@{}cc|cccc@{}}
%\toprule
$n$ & $FTBE_n$ & $N$ & $\omega$ & $\gamma_p$ & $\gamma_w$ \\ \midrule
200 & 0.072 & 1119 & 0.249 & 0.254 & 2.385\\
50 & 0.042 & 1117 & 0.250 & 0.264 & 3.572 \\
\bottomrule
\end{tabular}
\label{tbl:parameters}
\end{table}

{\color{black}\section{Roadmap to validation}\label{sec:roadmap}
Although our study is numerical, a number of recent experiments display pattern formation strikingly similar to the regimes we identify. For example, Quincke rollers confined in circular microchambers self-organize into steady boundary vortices punctuated by intermittent particle ejections into the interior~\cite{Bricard2015}, or into transient flock-like bands~\cite{Bricard2013-dg}. When confined to annular geometries containing obstacles, such systems can exhibit an oscillating global chirality~\cite{Zhang2020}, similar to our ordered oscillation pattern. 

Moving away from artificial to living systems, bacterial monolayers under confinement can develop counter‐rotating edge currents~\cite{Wioland2013}, suspected to be a generic phenomenon in bacterial suspensions with highly no-slip boundaries, e.g., oil interfaces. Confined to `racetrack'-like geometries they behave similarly to Quincke rollers; they form flock-like bands moving unidirectionally along the track~\cite{Wioland_2016}. 

These qualitative parallels suggest that the oscillating vortex regime maximising mixing in our simulations could be realised in bench‑top active matter platforms. To guide a concrete realisation, we envision experiments on pear-shaped polystyrene Quincke rollers in a circular microchamber~\cite{Zhang2020_2} subject to a perpendicular electric field, whose strength determines the particle self-propulsion speed $v_0$. The shape anisotropy of such particles imparts an intrinsic chirality while inelastic collisions---both roller-roller and roller-wall---have been shown to lead to the required alignment mechanisms, due to a build-up of spatial correlations~\cite{Bricard2013-dg,Hanke2013}. A soft, short-range repulsion may arise naturally from combined hydrodynamic and induced dipolar interactions~\cite{Imamura2023}. Together, these mechanisms account for every term in the equations of motion~\eqref{eq:EOM}, demonstrating that our model is fully amenable to experimental validation and that the optimised mixing phase can, in principle, be observed.}

\section{Conclusions}\label{sec:conclusions}
In this work, we have introduced a braid-topological framework to quantify mixing in a confined chiral Vicsek model, leveraging only simulated particle trajectory data. By mapping Lagrangian trajectories to algebraic braids, we computed two complementary measures---the rescaled writhe, capturing global stirring strength and direction, and the finite-time braiding exponent, serving as a proxy for topological entropy and mixing efficiency. Through a systematic exploration of parameter space (particle number, chirality, and alignment strengths), we demonstrated how distinct steady-state patterns---ranging from simple edge currents and single flocks to spiral droplets, travelling bands, and ordered oscillations---yield markedly different mixing efficiencies.

Our analysis also revealed an inverse relationship between writhe and FTBE: strongly biased rotations (high absolute writhe) produce low mixing, while patterns with low net circulation but complex local rearrangements (e.g., oscillatory motion, spiral droplets, travelling bands) maximize braid complexity. A Pareto-front analysis confirmed this trade-off between stirring and mixing. Finally, by directly maximizing the FTBE via a simplex optimizer, we identified an emergent ``oscillating-vortex" pattern---comprising a stable boundary vortex that periodically exchanges particles with an ordered oscillatory bulk---as the most effective mixer within the chiral Vicsek class.

Our results illustrate the power of topological methods for characterizing and optimizing mixing in active matter systems, particularly where traditional continuum measures fail. The braid-based approach requires only sparse trajectory data, making it well suited for experiments on microswimmers or biological suspensions. Moreover, the observed Pareto trade-off offers a design principle: to enhance mixing, one must moderate global rotation in favour of local, topologically nontrivial motion.

As a next step, it would be interesting to extend this framework to more complex geometries (e.g., non-circular confinements, porous media), incorporate hydrodynamic interactions, or study mixtures of active and passive tracers. Experimental validation using colloidal rollers or bacterial baths could further confirm the predictive power of braid measures. {\color{black} We provided such a design for an experimental realisation of our system---based on artificial rollers---which can serve as a roadmap for pinpointing the proposed optimal mixing pattern.} Beyond microfluidic mixing, the braiding framework established here may inform the design of reconfigurable active devices and our understanding of transport in natural and synthetic active systems, for instance through the determination of Lagrangian coherent structures acting as transport barriers~\cite{ALLSHOUSE2012,Yeung2020}.

\section*{Author contributions}
{\bf Y. W.:} Formal Analysis, Investigation, Software, Visualization. {\bf J. B.:} Conceptualization, Methodology, Project Administration, Supervision, Writing – Original Draft, Visualization.

\section*{Conflicts of interest}
There are no conflicts to declare.

\section*{Data availability}

Data for this article, including the code used for the generation of the research data, are available at Zenodo at https://doi.org/10.5281/zenodo.15309860.

\section*{Acknowledgements}
The authors thank K. Proesmans for enlightening discussions.  J.B. is supported by the Novo Nordisk Foundation with grant No. NNF18SA0035142.

%%%END OF MAIN TEXT%%%


\balance

%%%REFERENCES%%%
\bibliography{rsc}
\bibliographystyle{rsc}

\end{document}